\documentclass{sigchi-ext}
\usepackage[T1]{fontenc}
\usepackage{textcomp}
\usepackage[scaled=.92]{helvet} 
\usepackage{graphicx} 
\usepackage{balance}  
\usepackage{booktabs} 
\usepackage{ccicons}  
\usepackage{ragged2e} 

\def\plaintitle{Capturing the Practices, Challenges, and Needs of Transportation Decision-Makers} 

\def\emptyauthor{}
\def\plainkeywords{Decision-maker; decision-making; user study; persona; transportation management and planning.}

\title{\plaintitle}

\numberofauthors{6}
\author{%
    \alignauthor{
    \textbf{Nasim Sharbatdar}\\
    \textbf{Yassine Lamine}\\
    \affaddr{Dept. of Computer Engineering} \\
    \affaddr{Polytechnique Montr\'eal} \\
    \affaddr{Montr\'eal, Qu\'ebec Canada} \\
    \email{nasim.sharbatdar@polymtl.ca}\\
    \email{yassine.lamine@polymtl.ca} }
    \alignauthor{
    } \vfil 
    \alignauthor{
    \textbf{Brigitte Milord}\\
    \textbf{Catherine Morency}\\
    \affaddr{Dept. of Civil Engineering} \\
    \affaddr{Polytechnique Montr\'eal} \\
    \affaddr{Montr\'eal, Qu\'ebec Canada} \\
    \email{brigitte.milord@polymtl.ca}\\
    \email{catherine.morency@polymtl.ca}}
    \alignauthor{
    } \vfil 
    \alignauthor{
    \textbf{Jinghui Cheng}\\
    \affaddr{Dept. of Computer Engineering} \\
    \affaddr{Polytechnique Montr\'eal} \\
    \affaddr{Montr\'eal, Qu\'ebec Canada} \\
    \email{jinghui.cheng@polymtl.ca} }
}

\definecolor{linkColor}{RGB}{6,125,233}
\hypersetup{%
  pdftitle={\plaintitle},
  pdfauthor={\emptyauthor},
  pdfkeywords={\plainkeywords},
  bookmarksnumbered,
  pdfstartview={FitH},
  colorlinks,
  citecolor=black,
  filecolor=black,
  linkcolor=black,
  urlcolor=linkColor,
  breaklinks=true,
}

\begin{document}


\maketitle

\RaggedRight{} 

\begin{abstract}
  Transportation decision-makers from government agencies play an important role in addressing the traffic network conditions, which in turn, have a major impact on the well-being of citizens. The practices, challenges, and needs of this group of practitioners are less represented in the HCI literature. We address this gap through an interview study with 19 practitioners from Transports Qu\'ebec, a government agency responsible for transportation infrastructures in Qu\'ebec, Canada. We found that this group of decision-makers can most benefit from research about data analysis tools and platforms that (1) provide information to support data quality awareness, (2) are interoperable with other tools in the complex workflow of the practitioners, and (3) support intuitive and customizable visual analytics. These implications can also be informative to the design of tools supporting other decision-making tasks and domains.
\end{abstract}




\section{Introduction}
Traffic congestion is a matter that brings concerns in most populated cities, posing various economic, environmental, and social issues~\cite{SUN2014, INRIX2019, LaPresse2018}. 
Bad traffic conditions also cause accidents, increases air pollution, and result in energy waste. 
The traffic network conditions are largely determined by two interlocked human factors: (1) the behavior and needs of traffic network users (e.g. drivers and pedestrians) and (2) the perception and practices of transportation decision-makers. 

In HCI research, plenty of work has been done in understanding and supporting the first group of people (e.g. \cite{Balters2019, Obrist2013}), while the challenges and needs of the second group are less represented. We address this limitation and directly focus on understanding the practices, challenges, and needs of transportation decision-makers. Particularly, we report an interview study with practitioners from Montr\'eal working at Transports Qu\'ebec, a government agency responsible for transportation infrastructure in Qu\'ebec, Canada.

Montr\'eal is the second largest city in Canada, with a population of around 1.7 million. As many major cities, Montr\'eal experiences some severe traffic congestion, especially in the central business areas and in the road networks connecting the city with its suburbs. The traffic situation is also complicated by the frequent events that require traffic control and the severe winter weather conditions. The transportation management and planning practitioners often need to make difficult decisions, balancing short term traffic impacts and long term benefits and risks when planning road closures and road work.

We are particularly interested in understanding how this group of decision-makers used the traffic data to support their work. We also aimed at creating personas~\cite{Cooper1999} to summarize the challenges and needs of different types of transportation decision-makers and to communicate these elements with the larger community who focuses on creating tools to support transportation decision-making.

\subsection{Related Work}
Our work is most closely related to previous studies that focused on transportation planning support. For example, Masli et al. proposed a decision support tool that leveraged a community-driven, geographic wiki to assist transportation planners in route analysis~\cite{Masli2013}. Le Dantec et al. established the feasibility of using crowd-sourcing data collected from a smartphone app in supporting transportation planning~\cite{LeDantec2015}. Prouzeau et al. have also proposed a interactive wall display system that focused on comparing predictive traffic models with real traffic data for traffic situation exploration~\cite{Prouzeau2016}. Few previous work in this area have directly examined the perceptions and practices of transportation decision-makers. Our study addresses this gap.







\section{Methods}
We conducted interviews with 19 participants from Montr\'eal working at Transports Qu\'ebec who focus on various aspects of transportation management and planning.

\subsection{The participants}
Two employees of Transports Qu\'ebec supported us in recruiting the participants to cover diverse job functions in the agency. Our participants included 12 males and 7 females. Their professional experience ranged from two years to more than 30 years (median = 7 years). The participants focused on different aspects of transportation management and planning, including (1) transportation modeling, (2) planning and sustainable mobility, (3) road network design and traffic analysis, (4) managing construction projects focused on improving traffic conditions on main arterial roads. Table~\ref{tab:Participants} summarizes our participants.

\subsection{Interview method}
The interviews were conducted in December 2018; each session lasted approximately 45 minutes. During the interviews, we asked the participants about: (1) their roles in Transports Qu\'ebec and their main practice and expertise, (2) their needs in using traffic data for decision-making, (3) the challenges they experienced when using such data, and (4) their expectations of tools for analyzing traffic conditions on the road network. All interviews were conducted in French, the official language in Qu\'ebec. The interviews were recorded in audio and fully transcribed for analysis. The transcripts were later translated into English.

\begin{margintable}[1pc]
  \begin{minipage}{\marginparwidth}
    \centering
    \footnotesize
    \begin{tabular}{ll}
        ID & Expertise / Role\\
        \toprule
        P1 & Sustainable transport \\
        P2 & Road pavement management \\
        P3 & Road design \\
        P4 & Transport systems modeling \\
        P5 & Road pavement management \\
        P6 & Transport systems modeling \\
        P7 & Project delivery \\
        P8 & Sustainable transport \\
        P9 & Major bridge projects \\
        P10 & Transport planning \\
        P11 & Major tunnel projects \\
        P12 & Transport systems modeling \\
        P13 & Sustainable transport \\
        P14 & Transport systems modeling \\
        P15 & Transport systems modeling \\
        P16 & Road pavement management \\
        P17 & Road pavement management \\
        P18 & Transport systems modeling \\
        P19 & Sustainable transport \\
        \bottomrule
    \end{tabular}
    \caption{Participants' expertise and role.}~\label{tab:Participants}
  \end{minipage}
\end{margintable}

\subsection{Data analysis}

We adopted an inductive thematic analysis approach to analyze the interview data~\cite{Guest2011}. 
First, three researchers independently coded the interview data to identify initial ideas and themes. 
Then we carried out an affinity diagram activity to organize the codes in a hierarchy that reveals the common practices, challenges and needs of the participants in the context of using the traffic data for supporting decision-making. After reviewing our codes and transcripts during several meetings with our group, we finalized a coding schema illustrating an overview of our findings. 
Based on participants responses to our coding schema, we then categorized the participants and created four personas to concretize and communicate the different characteristics of transportation management and planning practitioners. 



\section{Themes identified}
We generated themes about the use of traffic data for decision-making in the following three main categories.

\subsection{Practice}
We considered participants' practices in using traffic data for decision-making as contexts for understanding their challenges and needs. Our participants collectively reported the use of four main types of data:
\vspace{-8pt}
\begin{itemize}[noitemsep,leftmargin=0.5cm]
	\item Traffic counts and flow, ideally classified by the type of vehicle (e.g. trucks vs cars);
	\item Travel time or speed data based on road sections;
	\item Travel demand among various origins and destinations;
	\item Contextual data, varied according to the tasks to be performed, including information about road closures, accidents, weather, lane geometry, the socio-economic profile of geographic areas and users, pedestrian and bicycle counts, land use and residential development, public transit usage, etc.
\end{itemize}
\vspace{-8pt}

Participants reported the use of a wide spectrum of recency and granularity of the data. At one end, tasks such as road work management mainly require recent or real-time traffic data with a fine temporal and spatial granularity (e.g. precise segments of the road in increments of 15 minutes). At the other end, tasks such as project planning require historical data with a more coarse temporal and spatial granularity (e.g. seasonal data of a region).

To collect, validate, and analyze this data, participants reported using a variety of tools. Some used more basic such as spreadsheets or Google Maps, while others applied complex ones such as MapInfo, SQL, or R. Participants also used advanced data analysis tools developed internally at Transports Qu\'ebec or externally by a third party. 
Using these tools, participants sometimes created 2D/3D graphics to observe the evolution of congestion over time and space. Some participants also produced daily reports and/or made visualizations to justify an intervention.

We also noted two main types of preferences for participants' choice of data and tools, associated with their job responsibilities. Several participants mentioned that they prefer to have access to the raw data to calculate the indicators they needed for analysis. This level of access gives them better control and better knowledge about the quality of the data that goes into their model. However, several others preferred tools that offer graphical analyses and pre-calculated indicators. For these people, traffic data are often secondary inputs to their tasks (e.g. a project manager who needs to quickly validate or transmit information).

\subsection{Challenges}
Participants have discussed several major challenges regarding data collection and use that are interesting to designers of decision-making tools.

Many participants mentioned the challenge concerning the availability of data. For example, usually only the most important road sections are equipped with permanent vehicle counting devices. Even for those sections, device fails and severe weather conditions can postpone data collection. 
For some practitioners, especially those who focus on a more coarse data granularity, the lack of data would not be a major issue. For others, especially those who focus on the management of road closures and construction planning, missing or lack of precision in data can cause unforeseen situations and problems.

The second most frequently discussed challenge is associated with the quality and transparency of the data. This is particularly problematic for data that come from the private sector, as P18 explains: ``\textit{I have a big concern for transparency. Especially with big data today. We are talking about Google data among others. It's a big black box. We do not know their quality [...] Often we have numbers, we use them. It's very dangerous. [...] If we have bad data, garbage in, garbage out, we will not have good models.}'' 

Several participants were also concerned about the difficulty of interpreting the data in the absence of contextual variables. For example, P17 mentioned: ``\textit{We validate the vehicle counting data, but sometimes we do not know why there has been a decline. It was observed, but we do not know why (...) is it because of the temperature, was there an accident somewhere, is it simply a seasonal decline?}''

Participants also explained that a lot of efforts had to be put into cleaning and validating the data. For example, P4 said: ``\textit{The data, like Bluetooth [for capturing traffic speed], contain a lot of noise. We must remove the noise. ... You have to remove the days that do not work.}'' Some participants mentioned having conducted long-term studies to collect and validate data, sometimes cross referencing data from other agencies. 

With regard to the analytical tools, having to move frequently from one platform to another has posed a major problem. For example, P10 said: ``\textit{The fact that you have to go from one application to another does not make it easy to work. True, we are going to get our information everywhere. [...] But I would prefer that everything be in one place in the end.}'' This challenge is originated from the fact that there is a lack of an integrated platform for collecting and analyzing traffic data; the interoperability of the current tools is not satisfiable (i.e. transferring data from one tool to another can be difficult, if not impossible).


\subsection{Needs}
With respect to the data, participants mentioned a number of needs that remain unfulfilled:
\vspace{-8pt}
\begin{itemize}[noitemsep,leftmargin=0.5cm]
    \item Having data independent of the private sector, so that it can be cross-checked;
    \item Having more complete historical data, mostly related to contextual variables that would allow better data interpretation;
    \item Information on the method of data collection, the sample sizes and the accuracy of the estimates;
    \item Real-time data for managing road closures or construction sites.
\end{itemize}
\vspace{-8pt}

The participants summarized the ideal tools for transportation decision-making as being easy to use, flexible and multi-functional. They should include contextual variables and allow importing/exporting to ensure interoperability with other tools. For example, P1 emphasized the importance of having data to differentiate congestion types (e.g. caused by accident or weather condition). The tools should also provide information on the quality, availability, and accuracy of the data. 

Data visualization is regarded as an important feature in such tools. Participants mentioned that they needed tools to visualize summary information to facilitate data analysis and to speed up reporting. For example, P11 mentioned: ``\textit{We often need to do repetitive things - that is, reports for the month before, the month after, the week, whatever ... So it's important to have visualization that is quick to do, attractive and informative.}'' Direct manipulation when analyzing and visualizing the data is also a common need, as P1 mentioned: ``\textit{What would be interesting is to have access to heatmap, for example, according to the hours of the day and different points on the network, with a color representation of the traffic conditions with precise speeds.}''



\section{The personas}
Informed by these themes, we created four personas that captured different roles of practitioners within the transportation decision-making agency: (1) \textbf{Mona}, the modeling expert, who focuses on the creation and analysis of transportation system models; (2) \textbf{Tom}, the transport planner, which focuses on the analysis of the impacts of network intervention requests; (3) \textbf{Sam}, the safety expert, who focuses on various aspects of road safety including signaling and speed limits; and (4) \textbf{Caroline}, the coordinator, who supervises various projects and conducts opportunity studies. An overview of the personas can be accessed at: \href{https://github.com/HCDLab/TDMPersonas}{https://github.com/HCDLab/TDMPersonas}.

\section{Lessons Learned and Takeaways}
Through this interview study, we identified several opportunities for designing data analysis and decision support tools in the context of transportation management and planning. We are currently in the process of creating such a tool. Because many of these takeaways concern tool support for understanding data quality and interpreting data, they may also be informative to the design of tools supporting other decision-making tasks and domains.

\subsection{Support quality-aware data analysis}
One of the main takeaways in our studies is that the traffic decision-makers frequently face the problem of determining the quality and reliability of their data. While their decision-making work requires an understanding of data quality, such information is usually not evident. This problem manifested in multiple aspects. The data gathered from third-party platforms usually do not contain information about data quality. The participants also frequently experience missing data, which resulted in varied data quality in temporal, spatial, and contextual dimensions. The tools they currently use also do not provide sufficient support for analyzing data quality and reliability. As a result, participants needed to spend a lot of extra effort to validate and clean their data. In situations where they are not or can not get informed about data quality, their decision-making process and outcomes will be largely affected.

This challenge can be addressed from several angles. First, quality-aware data analysis tools need to be developed to support the practitioners to understand the reliability of their decision-making outcomes. Particularly, techniques for representing uncertainty in data and analysis results need to be employed and adapted to support the work of this group of decision-makers. Second, updating and expanding the data collection infrastructure can partially address the missing data problem. Collaboration with private sectors (e.g. taxi companies) and research institutes can also increase the data availability. 

\subsection{Support interoperability among different analysis tools}
Our participants reported that they leveraged data from diverse sources in their work. As a result, they often had to use multiple tools to collect and analyze data. A big problem they experienced when using these tools is the lack of interoperability. In other words, it is difficult to switch among different tools, transferring the data and analysis results. Context loss is an important effect reported by our participants when they used multiple tools. Dependence on third-party tools and platforms exacerbated this problem. 

To alleviate this problem, tool makers need to better consider the workflow of the transportation decision-makers, supporting them to easily switch among tools. Additionally, contextual information created and used during data analysis also needs to be stored and be transferable. For example, tools can be designed to record the traces of the steps involved in data analysis that eventually lead to a certain conclusion. Such information can support transparency and communication about the decision-making process.



\subsection{Supporting quick and informative data visualization}
Our participants have mentioned the importance of data visualization in multiple contexts. Particularly, they valued both bite-sized visualizations to quickly explore the data and more sophisticated on-demand visualizations based on their immediate temporal, spatial, and contextual interests in the analysis. In addition to visual analytics purposes, the practitioners also required intuitive visualizations for communicating their analysis results with other stakeholders in the agency, as well as with the general public.

Designing a customizable visualization system, based on the tasks and roles of the different types of practitioners (e.g. based on the personas we created), would then be a useful and interesting next step. Because of the diverse tasks this group of practitioners performs with visualization aids and their unique characteristics, balancing flexibility and ease of use is an important factor. Providing an appropriate level (i.e. sufficient and not overwhelming) of control during visual analytics and visual representation would be an important and challenging future work. 


\section{Conclusion}
In this study, we interviewed 19 transportation management and planning practitioners from Transports Qu\'ebec in Montr\'eal to explore their practices, challenges, and needs in data collection and analysis for decision-making. We also created four personas based on the themes identified in the interviews to capture the characteristics of diverse roles of these practitioners. Through this study, we revealed several research opportunities from the perspective of HCI that can support the work of this less represented group of users. These implications can also be informative to other decision-making tasks.

\section{Acknowledgements}
We thank all our participants for their time and insightful comments. We greatly acknowledge the support of Transports Qu\'ebec and particularly Mr. Guy Canuel and Mr. Marc-Andr\'e Tessier for all the coordination and helps.

\balance{} 

\bibliographystyle{SIGCHI-Reference-Format}
\bibliography{References}

\end{document}